\documentclass[sigconf]{acmart}

\usepackage[shortcuts,acronym]{glossaries}

\AtBeginDocument{%
  \providecommand\BibTeX{{%
    \normalfont B\kern-0.5em{\scshape i\kern-0.25em b}\kern-0.8em\TeX}}}

\copyrightyear{2022} 
\acmYear{2022} 
\setcopyright{acmlicensed}\acmConference[ICSE-SEIP '22]{44nd International Conference on Software Engineering: Software Engineering in Practice}{May 21--29, 2022}{Pittsburgh, PA, USA}
\acmBooktitle{44nd International Conference on Software Engineering: Software Engineering in Practice (ICSE-SEIP '22), May 21--29, 2022, Pittsburgh, PA, USA}
\acmPrice{15.00}
\acmDOI{10.1145/3510457.3513065}
\acmISBN{978-1-4503-9226-6/22/05}

\begin{document}

\title{Using a Semantic Knowledge Base to Improve the Management of Security Reports in Industrial DevOps Projects}

\author{Markus Voggenreiter}
\email{Markus.Voggenreiter@siemens.com}
\orcid{0000-0003-3964-1983}
\affiliation{%
  \institution{Siemens Technology and LMU Munich}
  \city{Munich}
  \country{Germany}
}

\author{Ulrich Schöpp}
\email{schoepp@fortiss.org}
\affiliation{%
  \institution{fortiss GmbH}
  \city{Munich}
  \country{Germany}
}

\renewcommand{\shortauthors}{Voggenreiter and Schöpp}


\newacronym{sast}{SAST}{Static Application Security Tool}
\newacronym{dast}{DAST}{Dynamic Application Security Tool}
\newacronym{vulnscan}{VST}{3rd Party Vulnerability Scanning Tool}
\newacronym{kb}{KB}{Knowledge Base}

\begin{abstract}
    Integrating security activities into the software development lifecycle
to detect security flaws is essential for any project. These
activities produce reports that must be managed and looped back
to project stakeholders like developers to enable security improvements.
This so-called Feedback Loop is a crucial part of any project
and is required by various industrial security standards and models.

However, the operation of this loop presents a variety of challenges.
These challenges range from ensuring that feedback data is
of sufficient quality over providing different stakeholders with the
information they need to the enormous effort to manage the reports.
In this paper, we propose a novel approach for treating findings
from security activity reports as belief in a \ac{kb}.
By utilizing continuous logical inferences, we derive information
necessary for practitioners and address existing challenges in the
industry. This approach is currently evaluated in industrial DevOps
projects, using data from continuous security testing.
\end{abstract}



\maketitle

\section{Introduction}
Automating security activities like periodic testing for vulnerabilities and flaws is essential in industrial projects utilizing DevOps techniques \cite{Moyon2020, kim_top_2011} to produce software products in domains with high security-related demand. As a result, new data about the software security is continuously generated, informing about shortcomings of the software and new requirements. In order to improve the product, reports must be fed back into the development cycle and be addressed by developers \cite{simpson_safecode_2014}. This so-called Feedback Loop is demanded by various standards and industry best practices \cite{iec4_1,migues_bsimm_2020}.  

In practice, however, the task of implementing the Feedback Loop presents various challenges. 
The first challenge is the quality of the reports. The vast amount of data produced by security activities varies in format, content, perspective, assumptions, and evaluation \cite{Welberg2008}, which necessitates data processing. Issues like False Positives are common in reports \cite{Nadeem_2012}, reducing their reliability. 
The second challenge is how the data is utilized. The correct interpretation of the security activity data is essential for the subsequent actions, as data itself fuels project decisions and represents the software’s security level. Moreover, each project has its own demands, e.g., regarding standards compliance. Consequently, a high-quality demand applies to the produced information and must be customized project-wise. 
Our third challenge is that data from security activities is only half of what is needed. Inputs by security experts, customer opinion, or vulnerability databases are essential to correctly estimate the impact of findings and present a valid representation of reality. Finally, performing the management manually to collate actionable information is neither feasible in industrial software development projects nor conforming with the DevOps mindset, where automation of tasks is crucial \cite{Leite_2019}. Hence, we see a necessity to address the question:\\
\textbf{How can the Feedback Loop for security reports in industrial DevOps projects be optimized?}\\
The optimization should include the process being \textit{faster}, \textit{customizable}, with the \textit{least manual effort}, \textit{more automation}, \textit{highly reliable}, and \textit{comprehensible} for the project team.
\section{Using a Semantic Knowledge Base}
\subsection{Theory}
We propose the usage of a semantic \ac{kb} to address the challenges identified above. 
A \ac{kb} comprises \textit{primary information}, logically interconnected by \textit{database semantics} and stored free of constraints in a \textit{metastructural database}. In contrast to a database, a KB has \textit{primary methods of data processing}, which allow the continuous generation of new information based on existing data \cite{Krotkiewicz2018,Krotkiewicz2016}. 
\ac{kb}s have been applied in various areas, including management of sensor data~\cite{Nambi2014}, the elicitation of high-quality requirements~\cite{Kaiya2006}, and even vulnerability management~\cite{Wang2009}. In contrast to existing approaches, we apply \ac{kb}s to the domain of secure software engineering to manage security reports. 
However, this implies multiple changes to existing concepts to ensure a successful function in this use case. Initially, we consider the content of the \ac{kb} as belief instead of knowledge due to the lack of data reliability. This particularly includes contradictory information from our data sources. Consequently, belief must be revisable, including the explicit belief (provided from outside the \ac{kb}) and derived belief (derived by the \ac{kb}). Moreover, the inference procedure must be customized to each project. This implies that each \ac{kb} has to deal with belief and inference rules being incrementally added or changed throughout the project. As each \ac{kb} is unique to its project, a traditional approach using a static ontology is not feasible for our use case, especially when considering changing inference rules. Instead, we enable the incremental creation of a \ac{kb} for each project. To ensure a consistent \ac{kb}, we monitor all changes to the \ac{kb} and identify contradictory information. If conflicting data is identified, it is resolved by considering external human input as more reliable (e.g., False Positive identification). Any belief derived from the formerly contradictory information is revised and re-calculated.  

With this approach, we can address various challenges of the industrial security report management with a semantic \ac{kb}. The automated generation of data with pre-written inference rules ensures high reliability, high comprehensibility, project customization and reduces the scope for human errors. The automation of the report management reduces the manual effort and speeds up processing. 

\subsection{Practice}
\begin{figure}[]
    \includegraphics[trim={1.3cm 21.6cm 0.8cm 1.3cm},clip,width=\linewidth]{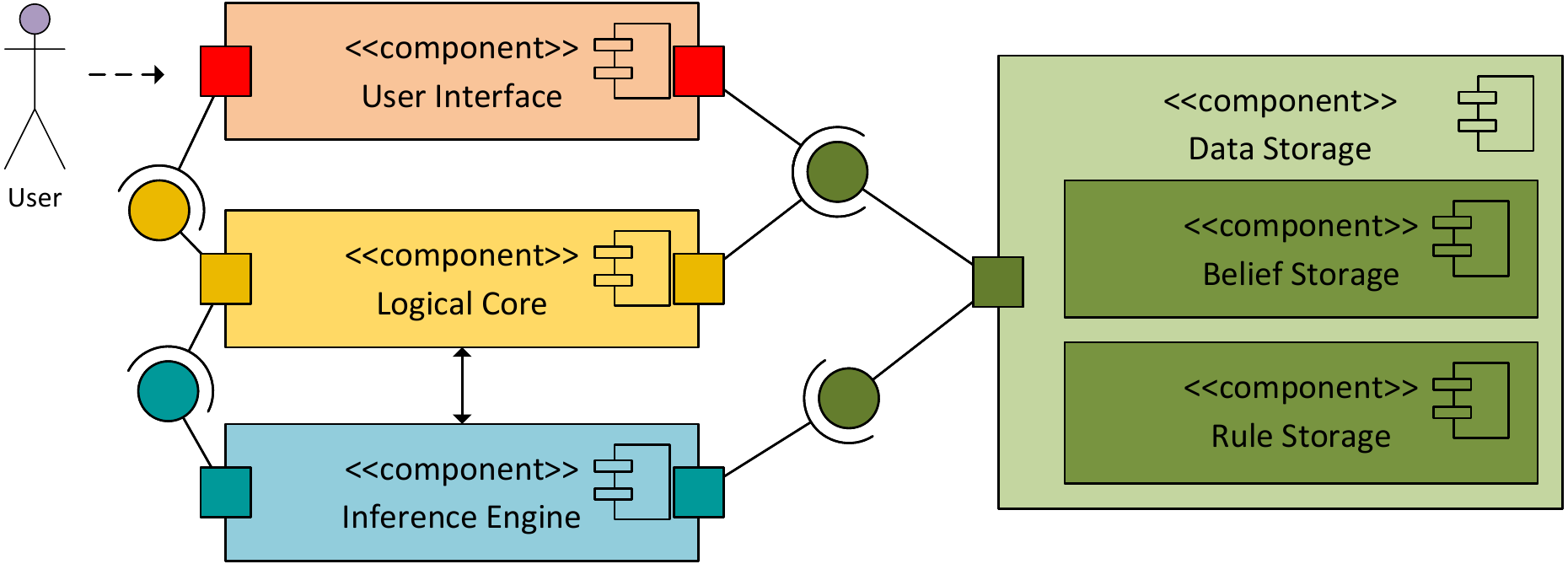} 
	\caption{Component Overview of Example \ac{kb}} 
	\label{fig:component_diagram} 
\end{figure}

In order to test our theoretical concept, we implemented it for industrial software development projects. We realized our concepts of belief, inference rules, continuous inference, and metastructural data storage in the components depicted in Figure~\ref{fig:component_diagram}.

The \textit{Data Storage} component, which comprises rules and belief, is implemented using the Elasticsearch search engine, which allows us to perform advanced queries on the data. Belief is stored in entities of Elasticsearch documents. Inference rules, however, are written in Python code. The step where our inference rules are applied to the current content of the \ac{kb} to derive new information is implemented by the \textit{Inference Engine}. With the \textit{Logical Core}, we ensure consistency between new beliefs being added or existing beliefs being revised. These components are implemented in Python, using a self-developed logic to ensure consistency within the \ac{kb}. Investigating the connections to incremental logic programming approaches, such as Differential Datalog \cite{differential_2022}, is an interesting direction for future work. 

In practice, each project has to customize the information contained in the \ac{kb} by writing documents for belief and inference rules. Based on our experience, specific inference rules are necessary for any project. These include a parser for security reports, a deduplication between findings, a validation of findings using human expertise, and a prioritization of the resulting issues. In most cases, these inference rules are streamlined, meaning that they build upon each other in a pipeline-like manner. To avoid the artificial inflation of the \ac{kb}, we restrict new inferences solely to those that might be invalidated by external input (e.g., incorrect deduplication). 

Deduplication, e.g., works on findings that have been parsed before from the original reports. During the execution of this rule, findings with similar values for title or description are summarized. As this could incorrectly connect two findings, external input correcting potentially flawed inferences is considered during execution. In such cases, the logical core would investigate whether this makes a revision of belief necessary. 
We are currently utilizing this \ac{kb} in industrial software development projects to manage reports from automated security testing to evaluate its relevance and usefulness.

\section{Conclusion and Future}
The management of reports produced by security activities in industrial DevOps projects is essential in every domain. In this paper, we suggest a novel approach of using a semantic KB for this use case. We indicate necessary changes to existing KB concepts and introduce our concept implementation. We believe that utilizing logical inferences in combination with the reliability of a KB is a promising approach for usage in industrial software engineering projects. Substantiating this belief with a long-term evaluation in a realistic setting will be the core activity of future work.


\bibliographystyle{ACM-Reference-Format}
\bibliography{sources}


\begin{thebibliography}{14}


\ifx \showCODEN    \undefined \def \showCODEN     #1{\unskip}     \fi
\ifx \showDOI      \undefined \def \showDOI       #1{#1}\fi
\ifx \showISBNx    \undefined \def \showISBNx     #1{\unskip}     \fi
\ifx \showISBNxiii \undefined \def \showISBNxiii  #1{\unskip}     \fi
\ifx \showISSN     \undefined \def \showISSN      #1{\unskip}     \fi
\ifx \showLCCN     \undefined \def \showLCCN      #1{\unskip}     \fi
\ifx \shownote     \undefined \def \shownote      #1{#1}          \fi
\ifx \showarticletitle \undefined \def \showarticletitle #1{#1}   \fi
\ifx \showURL      \undefined \def \showURL       {\relax}        \fi
\providecommand\bibfield[2]{#2}
\providecommand\bibinfo[2]{#2}
\providecommand\natexlab[1]{#1}
\providecommand\showeprint[2][]{arXiv:#2}

\bibitem[\protect\citeauthoryear{(IEC)}{(IEC)}{2018}]%
        {iec4_1}
\bibfield{author}{\bibinfo{person}{International Electrotechnical~Commission
  (IEC)}.} \bibinfo{year}{2018}\natexlab{}.
\newblock \bibinfo{booktitle}{\emph{62443-4-1}}.
\newblock \bibinfo{publisher}{{Security for industrial automation and control
  systems Part 4-1 Product security development life-cycle requirements}}.
\newblock
\showISBNx{978-2-8322-5239-0}


\bibitem[\protect\citeauthoryear{{Kaiya} and {Saeki}}{{Kaiya} and
  {Saeki}}{2006}]%
        {Kaiya2006}
\bibfield{author}{\bibinfo{person}{H. {Kaiya}} {and} \bibinfo{person}{M.
  {Saeki}}.} \bibinfo{year}{2006}\natexlab{}.
\newblock \showarticletitle{Using Domain Ontology as Domain Knowledge for
  Requirements Elicitation}. In \bibinfo{booktitle}{\emph{14th IEEE
  International Requirements Engineering Conference (RE'06)}}.
  \bibinfo{pages}{189--198}.
\newblock


\bibitem[\protect\citeauthoryear{Kim}{Kim}{2011}]%
        {kim_top_2011}
\bibfield{author}{\bibinfo{person}{Gene Kim}.} \bibinfo{year}{2011}\natexlab{}.
\newblock \bibinfo{booktitle}{\emph{Top 11 Things You Need to Know About
  {DevOps}}}.
\newblock


\bibitem[\protect\citeauthoryear{Kr{\'o}tkiewicz, Wojtkiewicz, and
  Jod{\l}owiec}{Kr{\'o}tkiewicz et~al\mbox{.}}{2018}]%
        {Krotkiewicz2018}
\bibfield{author}{\bibinfo{person}{Marek Kr{\'o}tkiewicz},
  \bibinfo{person}{Krystian Wojtkiewicz}, {and} \bibinfo{person}{Marcin
  Jod{\l}owiec}.} \bibinfo{year}{2018}\natexlab{}.
\newblock \showarticletitle{Towards Semantic Knowledge Base Definition}. In
  \bibinfo{booktitle}{\emph{Biomedical Engineering and Neuroscience}},
  \bibfield{editor}{\bibinfo{person}{Wojciech~P. Hunek} {and}
  \bibinfo{person}{Szczepan Paszkiel}} (Eds.). \bibinfo{publisher}{Springer
  International Publishing}, \bibinfo{address}{Cham},
  \bibinfo{pages}{218--239}.
\newblock
\showISBNx{978-3-319-75025-5}


\bibitem[\protect\citeauthoryear{Kr{\'o}tkiewicz, Wojtkiewicz, Jod{\l}owiec,
  and Pokuta}{Kr{\'o}tkiewicz et~al\mbox{.}}{2016}]%
        {Krotkiewicz2016}
\bibfield{author}{\bibinfo{person}{Marek Kr{\'o}tkiewicz},
  \bibinfo{person}{Krystian Wojtkiewicz}, \bibinfo{person}{Marcin
  Jod{\l}owiec}, {and} \bibinfo{person}{Waldemar Pokuta}.}
  \bibinfo{year}{2016}\natexlab{}.
\newblock \showarticletitle{Semantic Knowledge Base: Quantifiers and
  Multiplicity in Extended Semantic Networks Module}. In
  \bibinfo{booktitle}{\emph{Knowledge Engineering and Semantic Web}},
  \bibfield{editor}{\bibinfo{person}{Axel-Cyrille Ngonga~Ngomo} {and}
  \bibinfo{person}{Petr K{\v{r}}emen}} (Eds.). \bibinfo{publisher}{Springer
  International Publishing}, \bibinfo{address}{Cham},
  \bibinfo{pages}{173--187}.
\newblock
\showISBNx{978-3-319-45880-9}


\bibitem[\protect\citeauthoryear{Leite, Rocha, Kon, Milojicic, and
  Meirelles}{Leite et~al\mbox{.}}{2019}]%
        {Leite_2019}
\bibfield{author}{\bibinfo{person}{Leonardo Leite}, \bibinfo{person}{Carla
  Rocha}, \bibinfo{person}{Fabio Kon}, \bibinfo{person}{Dejan Milojicic}, {and}
  \bibinfo{person}{Paulo Meirelles}.} \bibinfo{year}{2019}\natexlab{}.
\newblock \showarticletitle{A Survey of DevOps Concepts and Challenges}.
\newblock \bibinfo{journal}{\emph{ACM Comput. Surv.}}  \bibinfo{volume}{52}
  (\bibinfo{date}{Nov.} \bibinfo{year}{2019}).
\newblock
\showISSN{0360-0300}


\bibitem[\protect\citeauthoryear{Migues, Steven, and Ware}{Migues
  et~al\mbox{.}}{2020}]%
        {migues_bsimm_2020}
\bibfield{author}{\bibinfo{person}{Sammy Migues}, \bibinfo{person}{John
  Steven}, {and} \bibinfo{person}{Mike Ware}.} \bibinfo{year}{2020}\natexlab{}.
\newblock \bibinfo{booktitle}{\emph{{BSIMM} 11}}.
\newblock


\bibitem[\protect\citeauthoryear{Moy{\'o}n, Soares, Pinto-Albuquerque, Mendez,
  and Beckers}{Moy{\'o}n et~al\mbox{.}}{2020}]%
        {Moyon2020}
\bibfield{author}{\bibinfo{person}{Fabiola Moy{\'o}n}, \bibinfo{person}{Rafael
  Soares}, \bibinfo{person}{Maria Pinto-Albuquerque}, \bibinfo{person}{Daniel
  Mendez}, {and} \bibinfo{person}{Kristian Beckers}.}
  \bibinfo{year}{2020}\natexlab{}.
\newblock \showarticletitle{Integration of Security Standards in DevOps
  Pipelines: An Industry Case Study}. In
  \bibinfo{booktitle}{\emph{Product-Focused Software Process Improvement}},
  \bibfield{editor}{\bibinfo{person}{Maurizio Morisio}, \bibinfo{person}{Marco
  Torchiano}, {and} \bibinfo{person}{Andreas Jedlitschka}} (Eds.).
  \bibinfo{publisher}{Springer International Publishing},
  \bibinfo{address}{Cham}, \bibinfo{pages}{434--452}.
\newblock
\showISBNx{978-3-030-64148-1}


\bibitem[\protect\citeauthoryear{Nadeem, Williams, and Allen}{Nadeem
  et~al\mbox{.}}{2012}]%
        {Nadeem_2012}
\bibfield{author}{\bibinfo{person}{Muhammad Nadeem}, \bibinfo{person}{Byron~J.
  Williams}, {and} \bibinfo{person}{Edward~B. Allen}.}
  \bibinfo{year}{2012}\natexlab{}.
\newblock \showarticletitle{High False Positive Detection of Security
  Vulnerabilities: A Case Study}. In \bibinfo{booktitle}{\emph{Proceedings of
  the 50th Annual Southeast Regional Conference}} (Tuscaloosa, Alabama)
  \emph{(\bibinfo{series}{ACM-SE '12})}. \bibinfo{publisher}{Association for
  Computing Machinery}, \bibinfo{address}{New York, NY, USA},
  \bibinfo{pages}{359–360}.
\newblock
\showISBNx{9781450312035}


\bibitem[\protect\citeauthoryear{{Nambi}, {Sarkar}, {Prasad}, and
  {Rahim}}{{Nambi} et~al\mbox{.}}{2014}]%
        {Nambi2014}
\bibfield{author}{\bibinfo{person}{S.~N. A.~U. {Nambi}}, \bibinfo{person}{C.
  {Sarkar}}, \bibinfo{person}{R.~V. {Prasad}}, {and} \bibinfo{person}{A.
  {Rahim}}.} \bibinfo{year}{2014}\natexlab{}.
\newblock \showarticletitle{A unified semantic knowledge base for IoT}. In
  \bibinfo{booktitle}{\emph{2014 IEEE World Forum on Internet of Things
  (WF-IoT)}}. \bibinfo{pages}{575--580}.
\newblock


\bibitem[\protect\citeauthoryear{Simpson}{Simpson}{2014}]%
        {simpson_safecode_2014}
\bibfield{author}{\bibinfo{person}{Stacy Simpson}.}
  \bibinfo{year}{2014}\natexlab{}.
\newblock \showarticletitle{{SAFECode} Whitepaper: Fundamental Practices for
  Secure Software Development 2nd Edition}. In \bibinfo{booktitle}{\emph{{ISSE}
  2014 Securing Electronic Business Processes}} (Wiesbaden),
  \bibfield{editor}{\bibinfo{person}{Helmut Reimer}, \bibinfo{person}{Norbert
  Pohlmann}, {and} \bibinfo{person}{Wolfgang Schneider}} (Eds.).
  \bibinfo{publisher}{Springer Fachmedien Wiesbaden}, \bibinfo{pages}{1--32}.
\newblock
\showISBNx{978-3-658-06708-3}


\bibitem[\protect\citeauthoryear{VMware}{VMware}{2022}]%
        {differential_2022}
\bibfield{author}{\bibinfo{person}{VMware}.} \bibinfo{year}{2022}\natexlab{}.
\newblock \bibinfo{booktitle}{\emph{Differential {Datalog} ({DDlog})}}.
\newblock
\urldef\tempurl%
\url{https://github.com/vmware/differential-datalog}
\showURL{%
\tempurl}
\newblock
\shownote{original-date: 2018-03-20.}


\bibitem[\protect\citeauthoryear{{Wang} and {Guo}}{{Wang} and {Guo}}{2009}]%
        {Wang2009}
\bibfield{author}{\bibinfo{person}{J.~A. {Wang}} {and} \bibinfo{person}{M.
  {Guo}}.} \bibinfo{year}{2009}\natexlab{}.
\newblock \showarticletitle{Security Data Mining in an Ontology for
  Vulnerability Management}. In \bibinfo{booktitle}{\emph{2009 International
  Joint Conference on Bioinformatics, Systems Biology and Intelligent
  Computing}}. \bibinfo{pages}{597--603}.
\newblock


\bibitem[\protect\citeauthoryear{Welberg}{Welberg}{2008}]%
        {Welberg2008}
\bibfield{author}{\bibinfo{person}{S.M. Welberg}.}
  \bibinfo{year}{2008}\natexlab{}.
\newblock \bibinfo{booktitle}{\emph{Vulnerability management tools for COTS
  software - A comparison}}.
\newblock Number TR-CTIT-08-15 in \bibinfo{series}{CTIT Technical Report
  Series}. \bibinfo{publisher}{Centre for Telematics and Information Technology
  (CTIT)}, \bibinfo{address}{Netherlands}.
\newblock


\end{thebibliography}

\end{document}